\documentclass[
   ,final            
  ]
  {aipproc}

\layoutstyle{6x9}

\begin{document}

\title{Spectroscopy and new particles}


\author{Tadeusz Lesiak
\footnote{partially supported by the KBN grant No. 2P03B 01324}}
{
  address={Institute of Nuclear Physics PAN \\
Radzikowskiego 152, 31-142 Krak\'{o}w, Poland}
}

\begin{abstract}
The properties of three new particles: X(3872), Y(3940) and Z(3931), recently
discovered by the Belle collaboration, are briefly reviewed.  
Negative results of the search for the pentaquark $\Theta(1540)^+$ are also presented.
\end{abstract}

\maketitle


\section{Introduction}

In the last two years the Belle collaboration
has provided evidence
for several new hadrons. This paper presents
the observation of three new particles: X(3872),
Y(3940) and Z(3931). The other new states
discovered recently by Belle were also
discussed at this conference in separate talks~\cite{KORPAR,TRABEL}.
The paper  also presents the results of searches
for the pentaquark state $\Theta(1540)^+$.
The Belle detector at the  KEKB asymmetric 
$e^+e^- $ collider~\cite{KEKB}
is a general purpose
spectrometer, described in detail in~\cite{BELLE}.

\section{Properties of the X(3872)}


\begin{figure}[bth]
  \includegraphics[height=3.7cm]{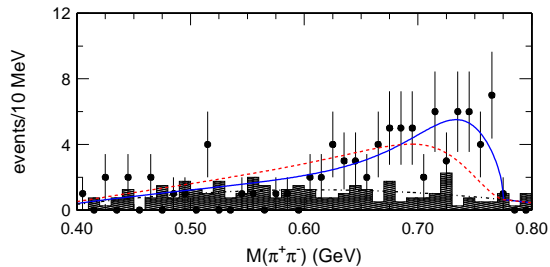}
  \caption{The distribution of $\pi^+\pi^-$ invariant mass for  events in the
           $X(3872)\to \pi^+\pi^- J/\psi$ signal region (data points). The shaded histogram corresponds
to the background as determined by the X-mass sidebands. The solid (dashed) curve
denotes a fit that uses a $\rho$ Breit-Wigner shape with the $\rho$ and $J/\psi$ in
a relative $S$($P$)-wave. The dash-dotted curve shows a smooth 
parametrization of the background that is used in the fit.}
\label{MASSPIPI}
\end{figure}


The state X(3872) was discovered by the Belle collaboration
in 2003~\cite{X1BELLE} by analyzing exclusive decays 
$B^+\rightarrow \pi^+\pi^-  J/\psi K^+$ (charge conjugate
modes are included everywhere, unless otherwise specified).
The $B$ mesons were reconstructed using two kinematical
variables: the energy offset $\Delta E = \sum_i E_i - E_{beam}$
and the beam-constrained
 mass $M_{bc}=\sqrt{E_{beam}^2 - \sum_i (\vec{p_i})^2}$,
where $E_i$ and $\vec{p_i}$  are the center-of-mass (CMS) energies and
momenta of the selected $B$ meson decay products and
$E_{beam}$ is the CMS beam energy.
A very narrow peak in the invariant
mass spectrum of the system $\pi^+\pi^- J/\psi$ was observed
with a mass of $3872.0\pm 0.6\pm 0.5$ MeV/c$^2$ and a width
below 2.3 MeV (90\% C.L.).

The observation of X(3872) was very quickly confirmed by the CDF~\cite{XCDF},
D0~\cite{XD0} and BaBar~\cite{XBABAR} experiments.
The observed decay mode $X\rightarrow \pi^+\pi^- J/\psi$ 
seemed to favour the explanation of the X(3872) as an excited 
charmonium state~\cite{EICHTEN1,EICHTEN2}. However, its properties, in particular
the very narrow width, did not allow the identification of the X(3872) with any $c\bar{c}$
state. At the same time the coincidence of the X mass with the 
$D^0\bar{D^{*0}}$ threshold ($3871.3\pm 1.0$)~MeV/c$^2$) has prompted many
theoretical speculations that X(3872) may be a so-called deuson i.e. 
a loosely bound molecular state of these two mesons~\cite{TORNQ,SWANSON}.
Moreover, the $\pi^+\pi^-$ invariant mass distribution (Fig.~\ref{MASSPIPI})
was found to  peak close to the upper kinematical limit of $M(\pi^+\pi^-)$
as expected for pion-pairs originating from $\rho\to\pi^+\pi^-$ decays.
\begin{figure}[b]
\begin{minipage}[b]{.5\linewidth}
\setlength{\unitlength}{1mm}
  \begin{picture}(70,40)
  \put(12,30){\bf a)}
    \includegraphics[height=3.9cm]{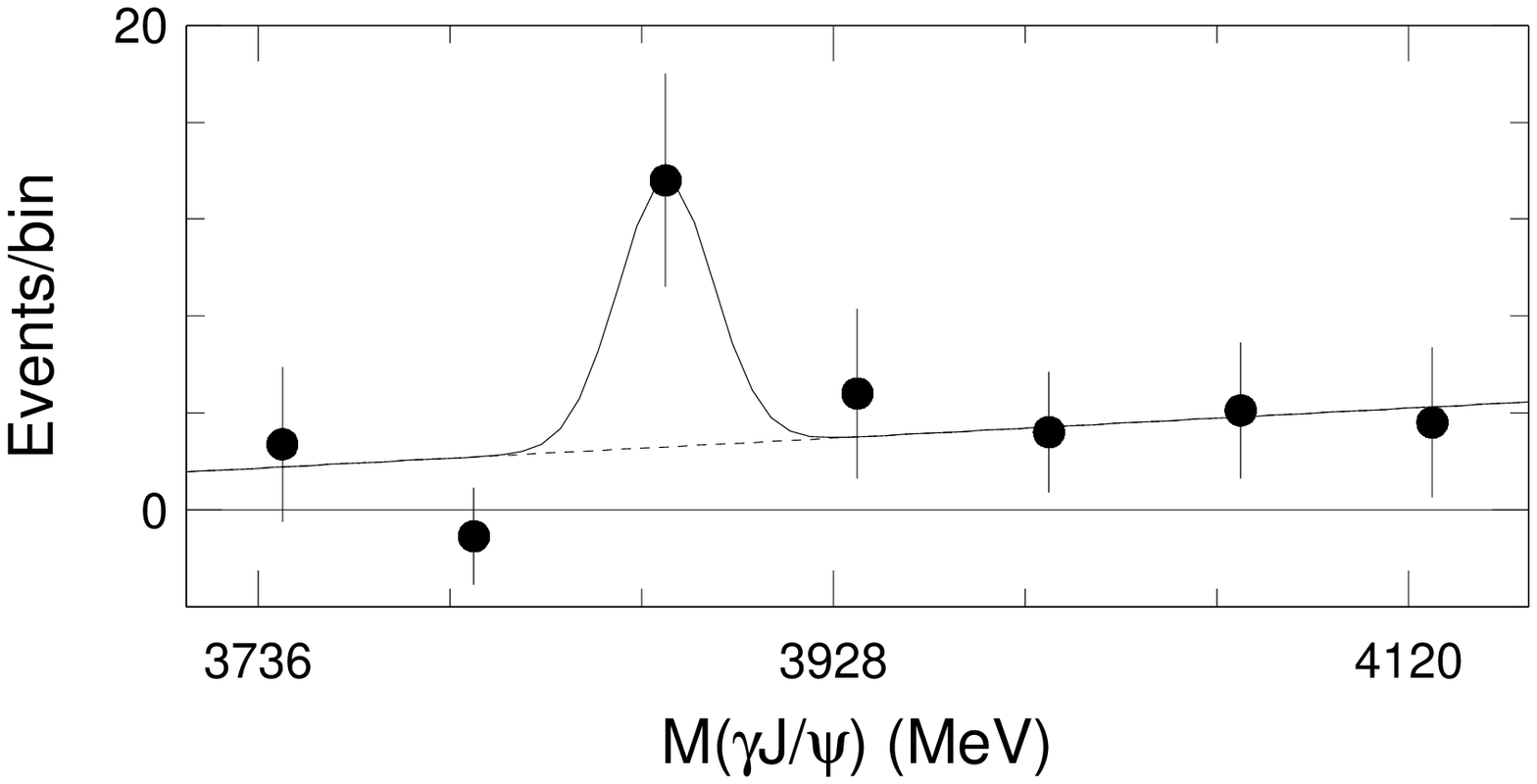}
  \end{picture}
\end{minipage}\hfill
\begin{minipage}[b]{.5\linewidth}
\setlength{\unitlength}{1mm}
  \begin{picture}(70,40)
  \put(12,30){\bf b)}
    \includegraphics[height=3.9cm]{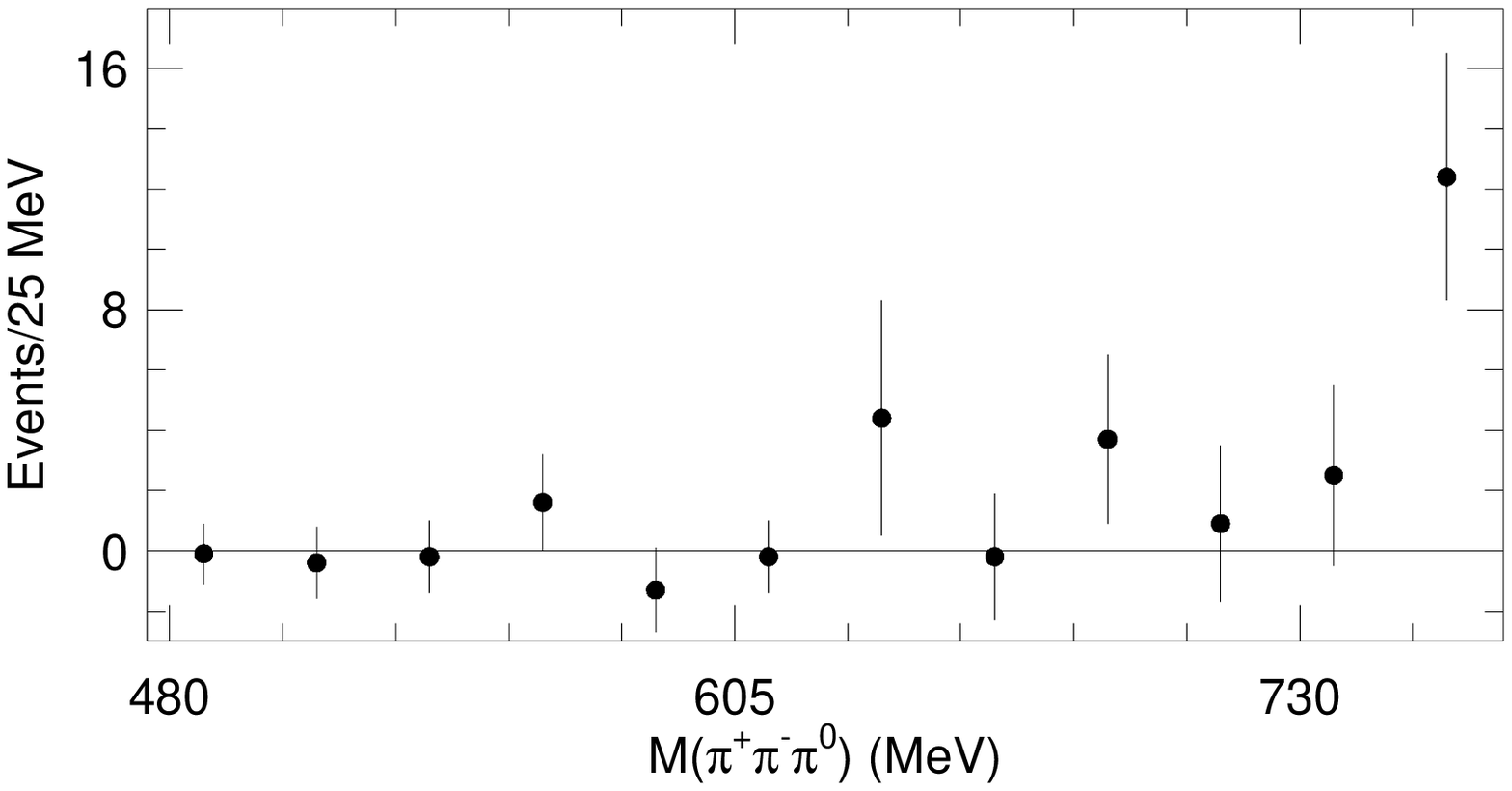}
  \end{picture}
\end{minipage}
  \caption{The yield of $B$ mesons from the decay {\bf a)}
             $B^0\rightarrow \gamma J/\psi K$,
            in bins of the  $\gamma J/\psi$ invariant mass and {\bf b)}
            $B^0\rightarrow \pi^+\pi^-\pi^0 J/\psi K$,
            in bins of the  $\pi^+\pi^-\pi^0$ invariant mass,
  determined from fits to the $\Delta E$ and $M_{bc}$ distributions.} 
\label{GAMM3PI}
\end{figure}


Recently, the Belle collaboration,  using 
the 253~fb$^{-1}$ data sample collected at the $\Upsilon(4S)$ resonance,
 has provided the first evidence for two
new decay modes: $X\rightarrow \gamma J/\psi$
and $X\rightarrow \pi^+\pi^-\pi^0 J/\psi$~\cite{BELLE540}.
They were observed in exclusive $B$ meson decays to the final states
$\gamma J/\psi K$ and $\pi^+\pi^-\pi^0 J/\psi K$, respectively.
The yield of the decay $B\rightarrow \gamma J/\psi K$ plotted in bins
of the $\gamma J/\psi$ invariant mass (Fig.~\ref{GAMM3PI}{\bf a)})
 exhibits an excess of 
$13.6\pm 4.4$ events (statistical significance of 4$\sigma$). 
The observation of this decay establishes unambiguously that the 
charge-conjugation parity of the X(3872) is positive.  
The partial width ratio 
$\Gamma(X\rightarrow \gamma J/\psi)/\Gamma(X\rightarrow \pi^+\pi^- J/\psi)$
amounts to $0.14\pm 0.05$. This result is in contradiction with the 
$\chi_{c1}^{\prime}$ ($1^{++}$ charmonium) assignment for X as in this 
case a value around 40 would be expected. The second decay mode 
$X\rightarrow \pi^+\pi^-\pi^0 J/\psi$ was found to be dominated by the
sub-threshold decay $X\rightarrow \omega^* J/\psi$. This is motivated by the
fact that the yield
of $B$ mesons plotted in bins of the $\pi^+\pi^-\pi^0$ invariant mass
(Fig~\ref{GAMM3PI}{\bf b)}) inside of the signal region from the decay
$X\rightarrow\pi^+\pi^-\pi^0 J/\psi$ is consistent with zero except
for the $M(\pi^+\pi^-\pi^0)>750$~MeV/c$^2$.
There, the excess of $12.4\pm 4.1$ events (4.3$\sigma$) is observed.
The ratio of branching fractions
{\cal B}($X\to\pi^+\pi^-\pi^0 J/\psi$)/{\cal B}($X\to\pi^+\pi^- J/\psi$).
was measured to be  $1.0\pm 0.4\pm 0.3$, which implies a large
violation of isospin symmetry.

The Belle collaboration also attempted  to determine the $J^{PC}$
quantum numbers of the X(3872)~\cite{BELLE541} by studying 
the angular distributions
of the decay $X\to\pi^+\pi^- J/\psi$, 
as suggested by Rosner~\cite{ROSNER}. Among the twelve possible $J^{PC}$
assignments, half
 ($0^{--}$, $0^{+-}$, $1^{--}$, $1^{+-}$, $2^{--}$ and $2^{+-}$) 
 may be discarded due to their negative charge conjugation-parity.
The value $1^{++}$ is in agreement with the data while
the assignments  $0^{-+}$  and $0^{++}$ are strongly disfavoured by the
analysis of angular distributions~\cite{BELLE541}.
The additional two odd-parity possibilities: 
$1^{-+}$ and $2^{-+}$  are discarded as for them the 
dipion invariant mass spectrum (Fig.~\ref{MASSPIPI}) is expected
to be much softer to compare with the data.
On the other side,
the distribution of $M(\pi^+\pi^+)$ is in agreement with the $1^{++}$
hypothesis.
The assignment $2^{++}$ was strongly disfavoured by the recent,
preliminary observation by Belle~\cite{BELLED0D0} of the decay
$B\to K X,~X\to D^0\overline{D^0}\pi^0 $, where the $11.3\pm 3.6$ 
 signal events (5.6$\sigma$) 
concentrate close to the threshold for the final state
$D^0\overline{D^0}\pi^0$. In  the $2^{++}$ case,
 the decay of a spin 2
state to  three pseudoscalars ($D^0\overline{D^0}\pi^0$) 
would require at least one pair of them to be in a relative D wave.
In such a configuration the near threshold production would be
strongly suppressed by a centrifugal barrier.

All the above observations strongly favour the assignment of $J^{PC}=1^{++}$
to the X(3872). This matches the expectations of 
models~\cite{TORNQ,SWANSON} interpreting the X as a $D^0\overline{D^{*0}}$
bound state. This hypothesis also explains the narrow width
of the X(3872) and  the shape of
$\pi^+\pi^-$ and $\pi^+\pi^-\pi^0$ spectra in its corresponding decays  and 
leads to the prediction of large isospin violation in the X decays.
Among alternative interpretations are: 
a `conventional' charmonium~\cite{EICHTEN1,EICHTEN2,XCCBAR},
glueball~\cite{SETH}, tetraquark~\cite{TETRAQ} or the 
so called cusp effect~\cite{CUSP}.

\begin{figure}[b]
  \includegraphics[height=4.0cm]{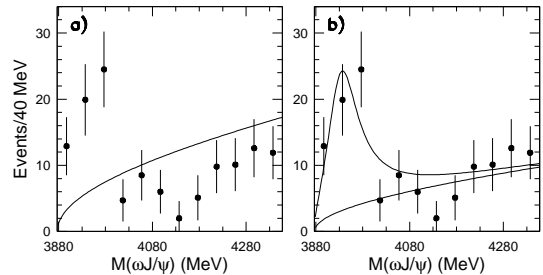}
  \caption{   $B^+\to K^+ \omega J/\psi$ signal yields vs  $M(\omega J/\psi)$. 
The curve in (a) shows the result of a fit that includes only a phase-space-like
threshold function. The curve in (b) corresponds to the result of a fit that
includes an $S$-wave Breit Wigner resonance term.} 
\label{YYYY}
\end{figure}


\section{Evidence for the Y(3940)}


In 2004 The Belle collaboration provided evidence for another
new state, Y(3940), decaying to $\omega J/\psi$\cite{Y3940}. It was again observed in
the $B^+$ meson exclusive decay to the final state $K^+ \pi^+\pi^-\pi^0 J/\psi$.
A fit to the $\omega J/\psi$ invariant-mass distribution (Fig.\ref{YYYY}) yielded
a signal of $58\pm 11$ events (8.1$\sigma$) corresponding to a mass of
 $3943\pm 11\pm 13$ MeV/c$^2$ and the width
$87\pm 22\pm 26$ MeV. The Y mass  coincides with that of another particle, X(3940),
also observed by Belle~\cite{KORPAR,TRABEL}. However, it is unlikely that these
two states are the same, since the X(3940) decays to  $D\bar{D^*}$ and does not decay
to  $\omega J/\psi$ and the situation is reversed for the Y(3940), as far as the
above-mentioned decays are concerned. The properties of Y(3940) are similar to those
expected for the $c\bar{c}-gluon$ hybrid mesons~\cite{HYBRID}.

\begin{figure}[t]
  \includegraphics[height=4.0cm]{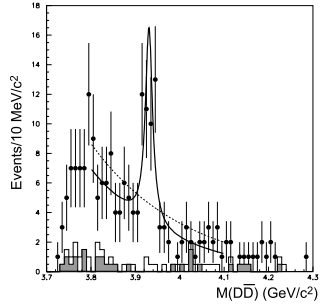}
  \caption{Invariant mass distribution of $D\overline{D}$ pairs. The solid (dashed) curve shows
the fits with (without) a resonance component. The histograms
correspond to the distribution of the events from the $D$-mass sidebands.} 
\label{ZZZZ}
\end{figure}

\section{Discovery of the Z(3931)}

A recent search by the Belle collaboration 
for the production of new resonances in the process $\gamma\gamma\to D\overline{D}$~\cite{BELLE576}
 yielded  evidence
for a new state (Fig.~\ref{ZZZZ}) at a mass of $3931\pm 4\pm 2$ MeV/c$^2$
and a width of $20\pm 8\pm 3$ MeV. A signal of $41\pm 11$ events with a
statistical significance of 5.5$\sigma$ was observed. The properties of this new state match the
expectations~\cite{ZPRIME,EICHTEN2} 
for the radially excited states
$\chi_{c0}^{\prime}$  and
$\chi_{c2}^{\prime}$. A study of angular distributions of the $D$ mesons in the $\gamma\gamma$
rest frame revealed that the data significantly prefer a spin two assignment over spin zero.


\begin{figure}[h]
  \includegraphics[height=5cm,width=7cm]{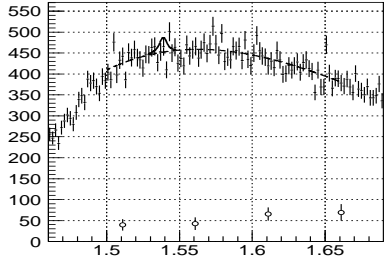}
  \caption{Invariant mass spectrum for secondary $p K^0_s$ pairs and expected yield
of the charge exchange reaction per 2 MeV/c$^2$ (open dots). A fit to a third order
polynomial is represented by the dashed curve. The $\Theta(1540)^+$ contribution expected
from the DIANA  result~\cite{DIANA} is presented with the solid line.} 
\label{PQPQ}
\end{figure}


\section{Search for the $\Theta(1540)^+$}


The Belle collaboration has searched for both inclusive and exclusive production of the
 $\Theta(1540)^+$ pentaquark using kaon secondary interactions in the material 
of the detector~\cite{BELLE518}. An upper limit of 2.5~\% (90~\% C.L.) was set on the
ratio of the $\Theta(1540)^+$ to $\Lambda(1520)$ inclusive production cross section.
The search for the exclusive production of the  $\Theta(1540)^+$ as an intermediate 
resonance in the charge exchange reaction $K^+n\to p K^0_s$ yielded an upper limit
of $\Gamma_{\Theta^+}< 0.64 $ MeV (90~\% C.L.) at $m_{\Theta^+} = 1.539$ MeV/c$^2$.
This value is below the current Particle Data Group~\cite{PDG} value of $0.9\pm 0.3$.


\section{Summary}


The properties of three new particles: X(3872), Y(3940) and Z(3931), recently
observed by the Belle collaboration,, were reviewed. For the X(3872) 
the observation of new decay modes together with angular analysis of the $\pi^+\pi^- J/\psi$ favours 
the assignment $J^{PC}=1^{++}$ and is in agreement with the deuson hypothesis.
The most plausible interpretations of the Y(3940) and Z(3931) are the 
$c\overline{c}-gluon$ and $\chi_{c2}^{\prime}$, respectively. The search for the 
$\Theta(1540)^+$ yielded a null result giving rise to the limit $\Gamma_{\Theta^+}< 0.64 $ MeV~(90~\% C.L.).


 

\bibliographystyle{aipproc}   

\end{document}